\documentclass[final]{IEEEtran}
\usepackage[T1]{fontenc}
\usepackage{amsmath}
\usepackage[cmintegrals]{newtxmath}
\usepackage{graphicx}
\usepackage{subcaption}
\usepackage{xcolor}
\usepackage[nocompress]{cite}

\begin{document}
\bstctlcite{BSTcontrol}  

\title{A Flexible IAB Architecture for Beyond 5G Network}
\author{\IEEEauthorblockN{Shashi~Ranjan\IEEEauthorrefmark{1}, Pranav~Jha\IEEEauthorrefmark{1}, Abhay~Karandikar\IEEEauthorrefmark{1}\IEEEauthorrefmark{2}, and Prasanna~Chaporkar\IEEEauthorrefmark{1} \\}
\IEEEauthorblockA{\IEEEauthorrefmark{1}Department of Electrical Engineering, Indian Institute of Technology Bombay, Mumbai, India\\ Email: \{shashi.svk, pranavjha, karandi, chaporakar\}@ee.iitb.ac.in \\}
\IEEEauthorblockA{\IEEEauthorrefmark{2}Director, Indian Institute of Technology, Kanpur, India \\ Email: karandi@iitk.ac.in}}

\maketitle

\begin{abstract}
IAB is an innovative wireless backhaul solution to provide cost-efficient deployment of small cells for successful 5G adoption. Besides, IAB can utilize the same spectrum for access and backhaul purposes. The 3GPP standardized IAB in Release 16 and would incorporate a few enhancements in the upcoming releases.  The 3GPP IAB architecture, however,  suffers from some limitations, such as it does not support mobile relays or dual-connectivity. This article presents a novel IAB architecture that addresses these limitations and is transparent to legacy operations of the 5G system. The architecture also supports multi-RAT coexistence where access and backhaul may belong to different RATs. These factors (and many others) enable operators to capitalize on the architecture for deploying IAB anywhere in a plug-and-play manner. We also show the merits of the architecture by evaluating its capacity and mobility robustness compared to the 3GPP architecture. Simulation results corroborate our design approach. Owing its robust design, the architecture can contend for standardization in B5G system.

\end{abstract}


\section{Introduction}
Multihop wireless relaying in a cellular network has been a long due necessity. The use of relaying has shown to be a promising deployment solution for extending coverage area and sometimes boosting the network capacity of a cellular network \cite{hoymann2012}. It becomes more crucial in 5G, where ultradense deployment of millimeter-wave (mmWave) cells is the key enabler to meet 5G requirements such as high spectral efficiency, low energy consumption, and low latency. Ultradense networks pose high operation and capital costs for network operators because each base station (BS) needs to connect to the 5G core (5GC) through fiber backhaul. Furthermore, wired backhaul may face deployment restrictions in many urban as well as hard-to-reach areas. Wireless backhauling helps deploy ultradense networks easily and quickly without incurring additional wired backhaul costs. 

It is appealing that the mmWave spectrum is also a prime candidate for wireless backhaul. It exhibits large bandwidth, high directional links, and almost noise-limited characteristics \cite{akdeniz2014millimeter}. Hence, operators can leverage a part of the spectrum to provide fiber-like reliability. By not employing a separate spectrum license for backhaul, we can further reduce radio hardware and deployment costs. The above merits allow operators to envision diverse deployment scenarios in cellular networks, such as outdoor-to-indoor, outdoor small-cell, and group mobility (e.g., cells on buses or trains). 

Recognizing its importance to facilitate faster 5G deployments, the 3rd Generation Partnership Project (3GPP) standardized a solution for multihop relaying support over 5G new radio (NR) in Release 16 \cite{ts38300} called integrated access and backhaul (IAB). This feature supports both out-band and in-band backhauling. For the latter case, access and backhaul links can share the same spectrum in either time, frequency, or spatial domain. Many research activities have demonstrated the feasibility and potential of mmWave-based IAB networks using either end-to-end simulations \cite{polese2020integrated}, field trials \cite{tian2019field, cudak2021integrated}, or investigating routing and resource scheduling schemes \cite{saha2019millimeter, li2017radio}.

\begin{figure}[h]
\centering
 \includegraphics[width=\linewidth]{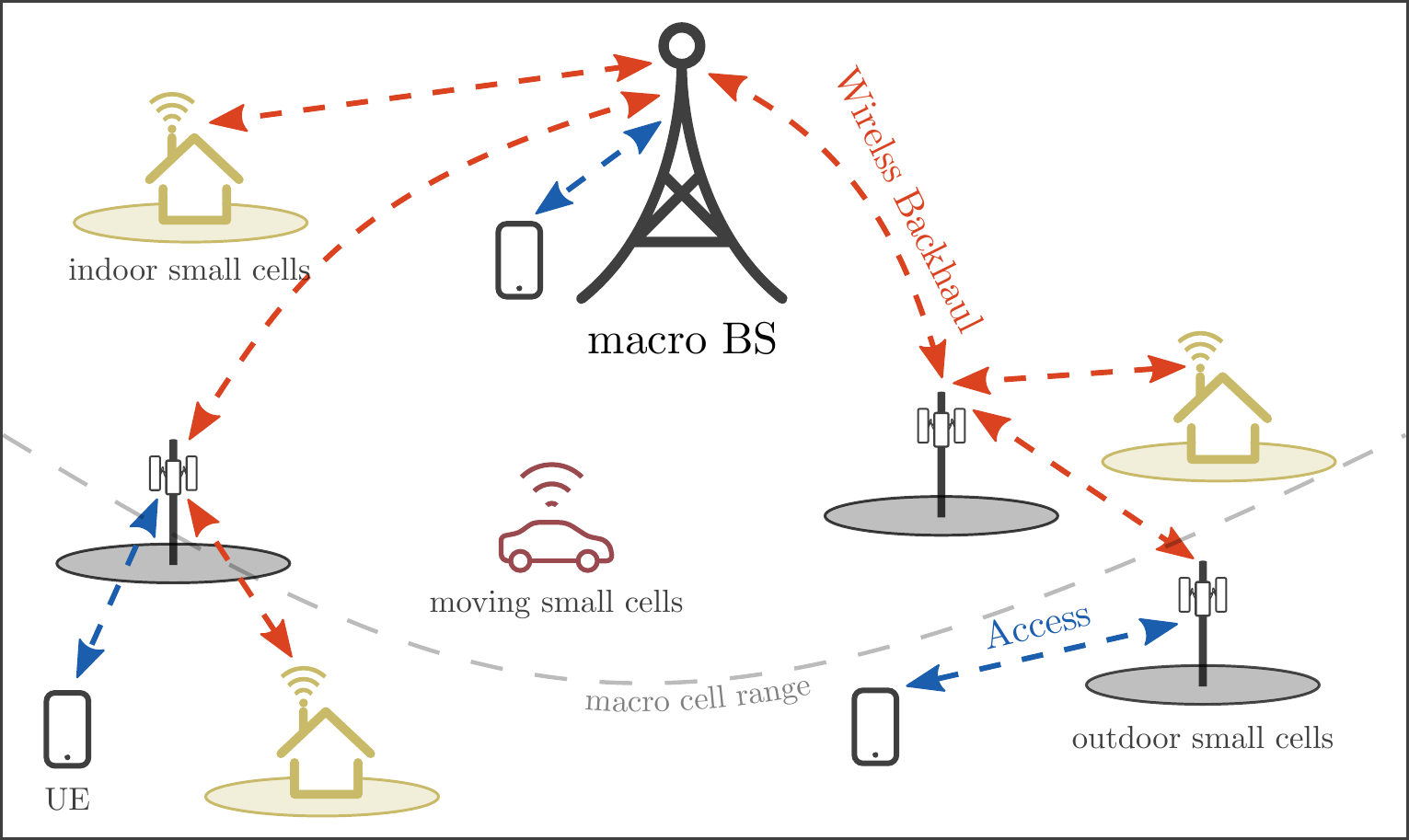}
\caption{An illustration of a cellular network structure that supports multihop relaying. Each relay (fixed or moving) creates its own small cell and the macro BS act as the anchor point for these cells.}
 \label{fig:illustration}
\end{figure}

An IAB network has two types of BSs: IAB-donors are the BSs that have fiber (or wired) backhaul to the 5GC, while IAB-nodes are the relays that use one or more wireless backhaul links to reach  IAB-donors. The IAB-donors thus act as access gateways for IAB-nodes and provide core connectivity to them. An IAB-node, therefore, plays a dual role -- as a user equipment (UE) from the perspective of upstream BSs and as a gNB (base station in NR) from the perspective of downstream BSs and UEs. Figure \ref{fig:illustration} represents an example of a multihop IAB network where macro BS is the IAB-donor, and small cell BSs are IAB-nodes.

The 3GPP 5G IAB architecture, however, has certain limitations that we will discuss later. This article aims to take a step in this direction by presenting a flexible IAB architecture for beyond 5G (B5G) networks. As we would observe later, the proposed architecture is flexible enough to handle almost every IAB design requirement and relaying deployments scenarios. Although the 3GPP is currently discussing some of them, the timeline for others is unknown at this point. The proposed architecture has minimal impact on the existing solutions for orchestration and management (OAM) and network service provisioning as 5GC is untouched. In essence, we design an IAB architecture that would help operators to incremental deploy 5G cells or extend an already deployed cellular network anytime, anywhere without worrying much about geographical restrictions and spectrum crunch.

\section{IAB Use Cases and Architecture Design Requirements} \label{sec:iab_requirements}
To enable the IAB feature, one needs modifications to the standard cellular system. These modifications can be, for example, a fresh design of the protocol stack or addition of new layers into the existing protocol stack to realize backhaul-related signaling and management. The 3GPP opted for the latter approach, which is more convenient and plausible. We may need to (re)design interfaces among IAB-nodes and between IAB-nodes and IAB-donors to carry signaling across multihop wireless links. We may also have to rethink C- and U-plane procedures such as bearer and mobility management, handover, and radio link failure (RLF) recovery. Incorporating these aspects transparently into the existing 5G radio access network (RAN) would achieve a robust IAB architecture. 

The architecture should have provisions to enable a smooth transition and flexible integration from/to the legacy deployments. By designing an architecture for legacy terminals, relaying operations become transparent to procedures for UEs, gNBs, and 5GC elements. Enforcing quality of service (QoS) for each bearer across IAB RAN is another facet that is paramount to architecture success. Since mmWave backhaul links are subject to blocking from moving objects or infrastructure changes, we also need topology adaptation and multi-connectivity capabilities. Under these features, it is possible to add or remove IAB-nodes to the network topology autonomously, recover from backhaul link overload/failure and reconfigure backhaul under local congestion in the event of variable traffic conditions. 

Support for mobile relays or vehicle mounted relays (VMRs) is necessary to improve the user experience for the onboard passengers in moving vehicles like buses or trains. Another merit of mobile relaying is group mobility, which prevents enormous signaling that may arise due to concurrent handovers of in-vehicle UEs. Mobile relay can also facilitate new deployment opportunities. For example, VMRs can be used to reach UEs with no or poor cellular coverage. Or vehicle-to-vehicle relaying can form a dynamic network and provide access to vehicles that become out of macro coverage (e.g., due to blocking) \cite{noh2020}. 

Lastly, NR coexistence with other radio access technologies (RATs) would allow operators to achieve better spectrum efficiency and wider coverage at lower costs. More importantly, in early deployments where NR might not have full coverage, the inter-RAT mobility between LTE and NR will often occur. With multi-RAT coexistence, IAB RAN can be deployed in areas with partial or no core connectivity by utilizing other RATs (e.g., LTE or wireless LAN (WLAN)) as overlay networks. Such a capability for IAB may give rise to newer deployment scenarios in the future. Both network and users would also benefit from faster mobility, enhanced service continuity, and augmented capacity when relays can maintain simultaneous connections to multiple cells belonging to different RATs. By aggregating these design aspects, we obtain a blueprint of an IAB network that can be deployed on sites where mobile traffic demand may arise.

\section{A Primer on 3GPP IAB Architecture} \label{sec:3gpp_arch}
The Release $16$ IAB specification defines radio protocols, upper layer management, and physical layer aspects for the 3GPP IAB architecture. The architecture has been designed to have low processing and design complexity at an IAB-node and minimal impact on the 5GC (and its related signaling overhead). Each IAB-node hosts a distributed unit (DU) and a UE part called mobile termination (MT). It is responsible for the lower layers of the radio interface to UEs or downstream MTs (of other IAB-nodes). The IAB-donor hosts a DU and the centralized unit (CU) that handles the control and upper layers of the radio interfaces. The IAB-donor-CU is responsible for IAB network topology and route management. It also manages radio resources if they are supervised centrally. Before an IAB-node can provide DU functionality, its MT registers to the 5GC and establish an NR connection with its parent node. Afterward, each IAB-node-DU maintains a logical F1 interface with the IAB-donor-CU regardless of its hop level. An additional layer called the backhaul adaptation protocol (BAP) layer on top of the radio link control (RLC) layer carries information on network topology and routing between the IAB-donor-CU and IAB-node-DUs. Specifically, the IAB-donor-CU configures BAP routing ID (carried in BAP header) on each IAB-node so that the BAP layer can route IP packets to the appropriate next node. Release 17, which may be available by March $2022$, is working on a few IAB enhancements such as improving load balancing, spectral efficiency, and multihop latency. The specifications \cite{ts38300, tr38874, ts38340} offer more details to interested readers on the architecture and its working.  

\begin{figure*}
\begin{minipage}{0.39884719156\textwidth}
\begin{subfigure}{\linewidth}
\includegraphics[width=\linewidth]{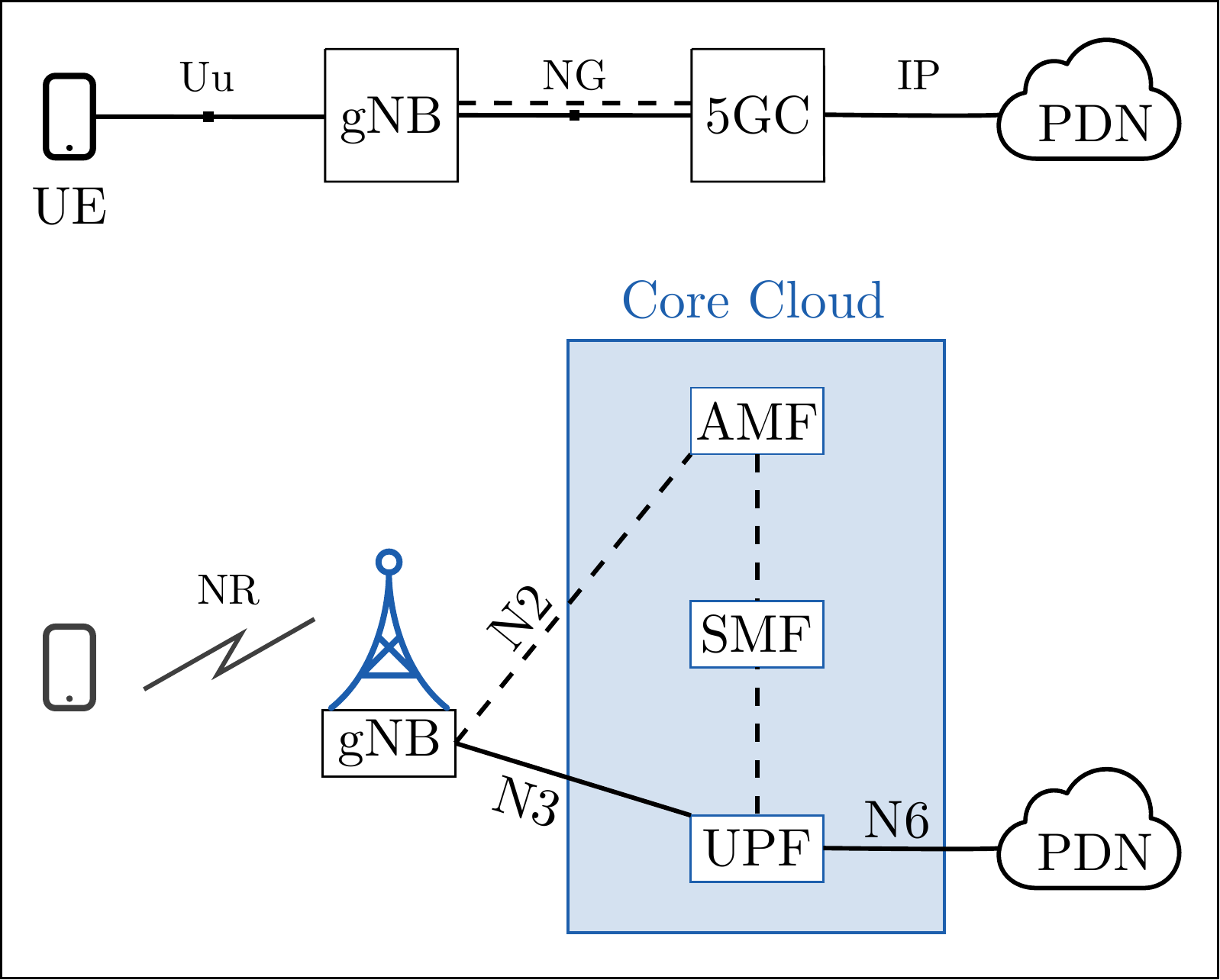}
\caption{} \label{subfig:part1}
\end{subfigure}
\vfill
\begin{subfigure}{\linewidth}
\includegraphics[width=\linewidth]{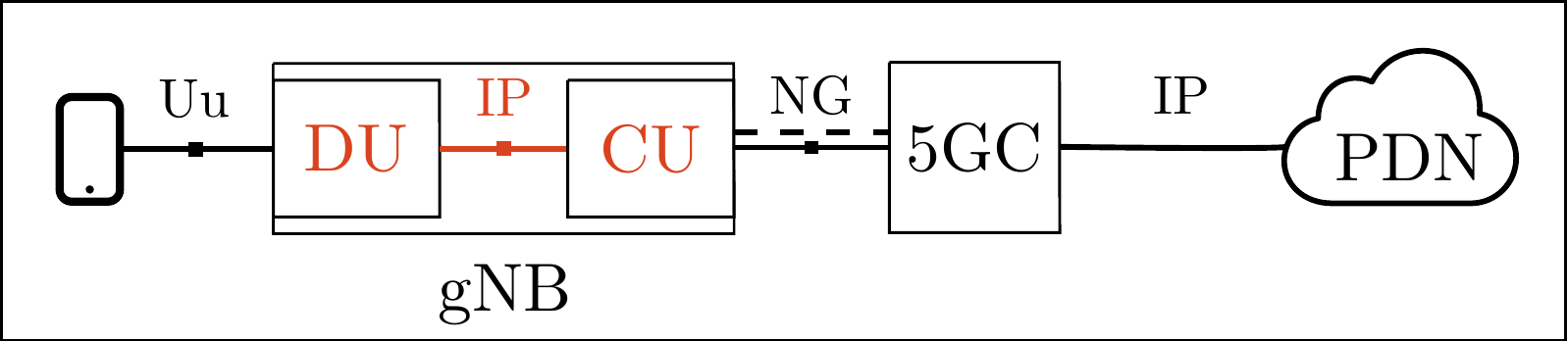}
\caption{} \label{subfig:part2}
\end{subfigure}
\end{minipage}
\hfill
\begin{subfigure}{0.60115280843\textwidth}
\includegraphics[width=\linewidth]{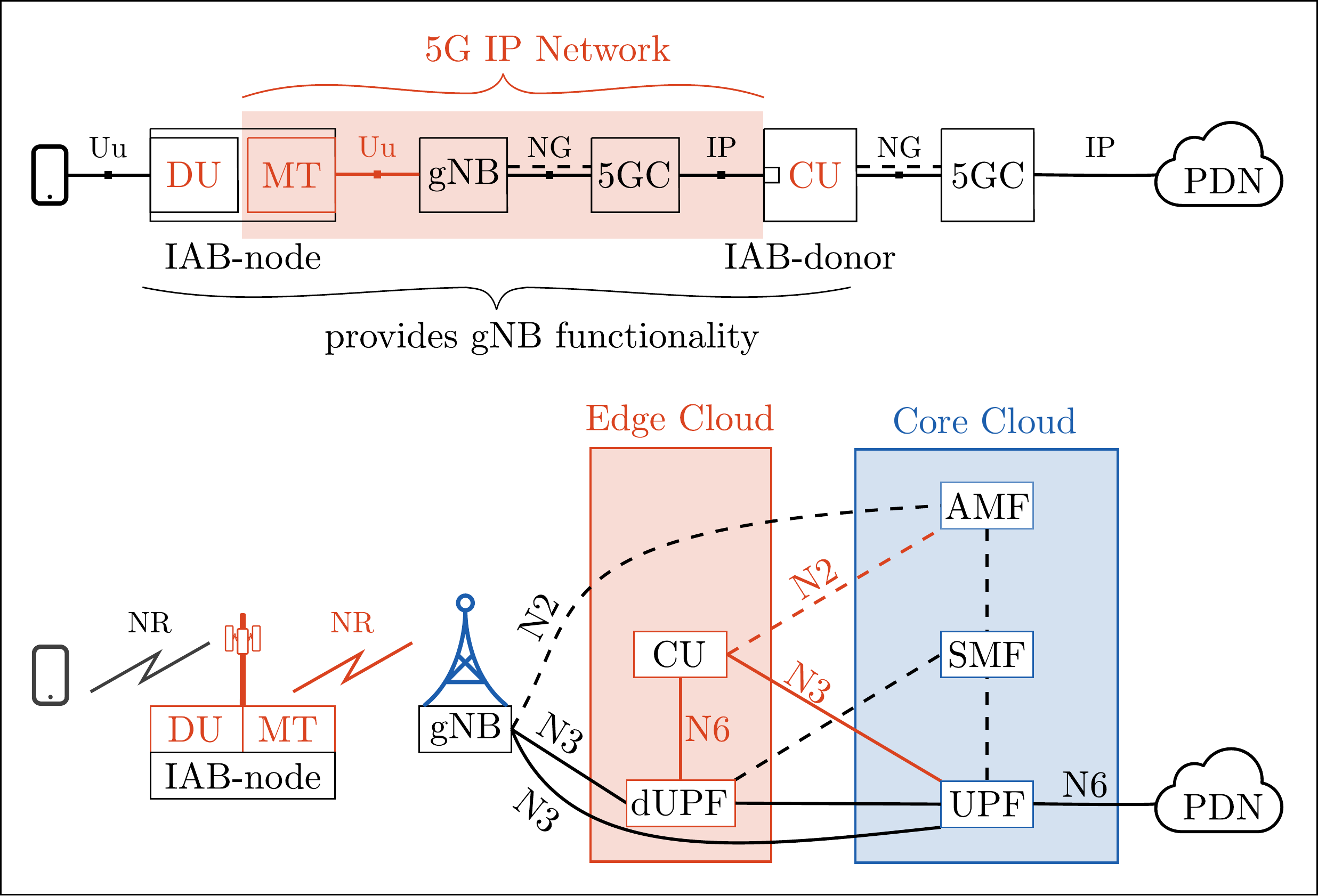}
\caption{} \label{subfig:part3}
\end{subfigure}
\caption{Evolution of the proposed IAB architecture from the standard 5G architecture along with network elements and their interfaces. The additional (IAB related) nodes and interfaces are in orange color. (a) Abstract view of the standard 3GPP 5G architecture. (b) Splitting of gNB as gNB-DU and gNB-CU units. Even if these units are not colocated but are connected via an IP network, together, they still provide gNB functionality. (c) Abstract view of the proposed architecture. The two units communicate via the 5G IP network to which they belong.}
\label{fig:arch}
\end{figure*}

That said, the 3GPP IAB architecture has a few limitations that may hinder its widespread commercial deployments. First, the architecture only supports fixed relays; IAB-nodes cannot move seamlessly across different parent nodes while retaining contexts of downstream MTs and UEs. Although migration of an IAB-node to another parent node 
underneath the same IAB-donor-CU is supported but only in the events of backhaul RLF recovery. Second, operators must deploy IAB-donors in a planned way before installing any IAB-node for proper relay operations. IAB-nodes can only work with and anchor through IAB-donors (that have special gNB-DUs to support relaying) and not with the traditional gNBs. This defeats the purpose of extending network coverage and deploying 5G cells flexibly wherever needed. The possibility of dual- or multi-connectivity for IAB-nodes also reduces because IAB-donors typically would have sparser deployment than gNBs. Third, the BAP layer requires a meticulous design for related C- and U-plane signaling. For example, we need enhancements in the radio resource control (RRC) layer to provide BAP addresses to IAB-nodes and treat an MT and UE differently at layer-2 and layer-3. Finally, the architecture does not support multi-RAT coexistence yet. As an example, we cannot use different RATs on access and backhaul at an IAB-node (e.g., LTE on access and NR on backhaul). This feature would act as a perfect platform for multi-RAT convergence in B5G networks where a UE can use any RAT for access, independent of RAT used for backhaul and associated core network.

As part of future releases, the 3GPP may work on new use cases and possibly mitigate some of these limitations. Some design issues, however, would remain the same. For example, even if Release $18$ supports mobile IAB, the movement would still be restricted to deployment regions having IAB-donors. 

\section{Proposed IAB Architecture}
We provide an alternate architecture to support multihop relaying by connecting gNB-CU and gNB-DU over the IP connection provided by the 5G network itself. Figure \ref{fig:arch} contains a graphic summary of how our IAB architecture has evolved from the standard 5G RAN architecture. To maintain consistency with the 3GPP, we have kept the same nomenclature for MT, IAB-node, and IAB-donor. The noticeable difference is that IAB-donor is now moved to the edge cloud and hosts the gNB-CU and a dedicated user plane function (dUPF). The latter forwards RLC and lower layers packets from IAB-node-DUs to the IAB-node-CU through its IP plane. Accordingly, the IAB-donor-CU acts as a server and is connected to the dUPF over the N$6$ interface. The IAB-node-DU and IAB-donor-CU communicate via the logical F1 interface over the 5G IP connection. The edge cloud is in the proximity of RAN, ensuring that C- and U-plane delays between MT and IAB-donor are affordable. We only need a single IAB-donor-CU at the edge cloud for all IAB-node-DUs within the network. We can place multiple IAB-donor-CUs as well if required for maintaining load balancing (or backup) and network robustness. In contrast, the 3GPP solution requires relatively more IAB-donors in the IAB RAN. Note that the logical separation of a gNB into two units is only for visualization purposes and understanding the architecture workings. In reality, the gNB (as shown in Fig. \ref{subfig:part3}) is indeed a BS and can exist as a single logical unit (i.e., a full gNB) or composition of CU and DU(s) apart from the IAB RAN elements.

\begin{figure*}
 \centering
 \begin{subfigure}{0.35779\textwidth}
\includegraphics[width=\linewidth]{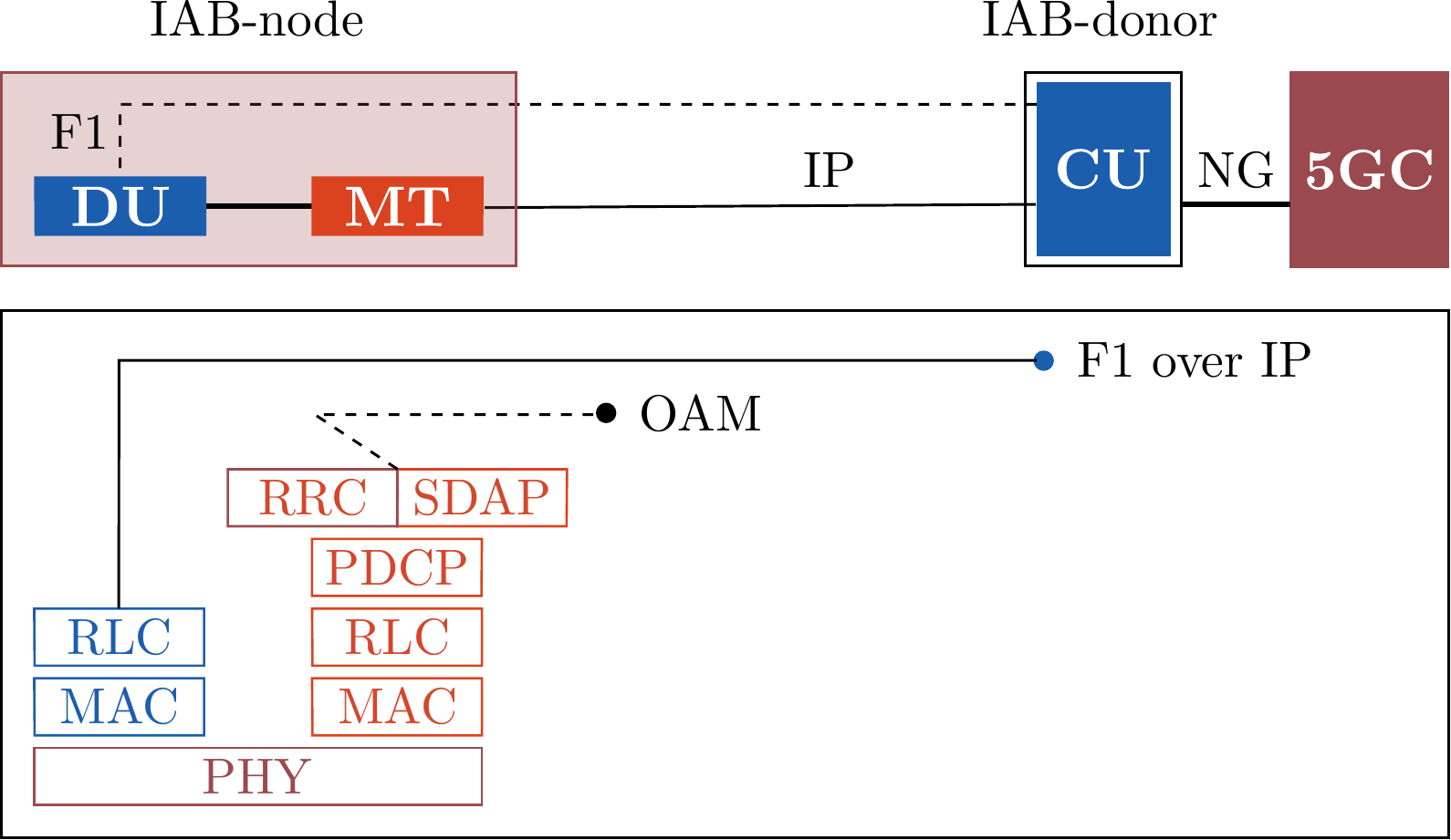}
\caption{} \label{subfig:two_stacks}
\end{subfigure}
\hfill
\begin{subfigure}{0.54221\textwidth}
\includegraphics[width=\linewidth]{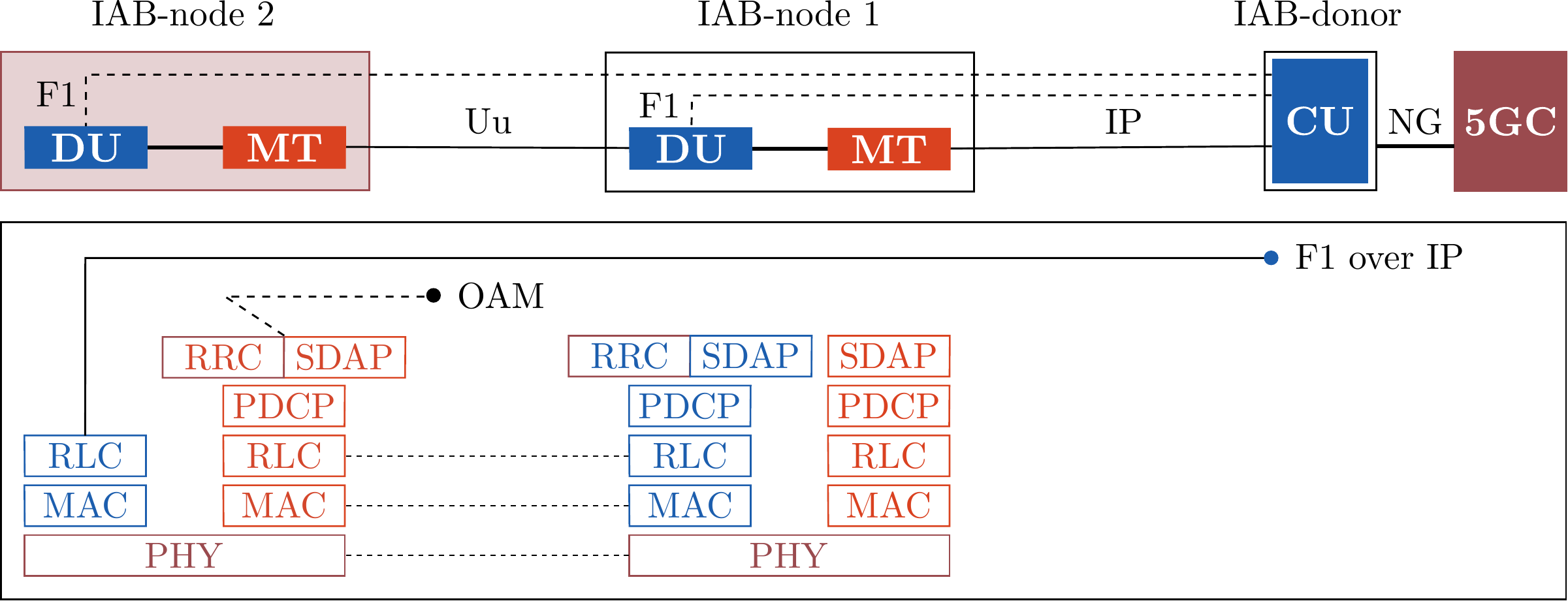}
\caption{} \label{subfig:multi_stacks}
\end{subfigure}
 \caption{Protocol stack at IAB-nodes in the proposed architecture for (a) two-hop and (b) multihop IAB system.}
 \label{fig:stacks}
\end{figure*}

The MT makes an NR connection to a gNB and reaches IAB-donor over the 5G IP network. The IAB-node-DU and the IAB-donor-CU then, in conjunction, provide gNB functionalities to prospective UEs. Therefore, the IAB-node forwards/receives C- and U-plane packets of UE to/from the 5GC. These packets are exchanged between the IAB-node-DU and the IAB-donor-CU through the F1 interface over the 5G IP connection established earlier. The IAB-donor-CU then communicates the packets with the 5GC over its N2/N3 interface. In the end, the UE packets follow the path IAB-node$\rightarrow$gNB$\rightarrow$dUPF$\rightarrow$IAB-donor-CU$\rightarrow$5GC.

The U-plane path of MT terminates at the IAB-donor-CU irrespective of its hop level. The MT obtains the IP address of its endpoint (i.e., IAB-donor-CU) from the OAM server during the initial setup procedure. If multiple IAB-donor-CUs are available, the OAM server can select the best one based on criteria like load situation and quality of backhaul links. An IAB-node, acting as an IP router for the previous (child or parent) node, forwards packets based on their destination IP addresses and selects appropriate RLC channels. Because of this flat IP design, the IAB network can support IP-related functionality efficiently, such as multicast, multipath, differentiated services, and well-known routing protocols. Many of these functions are taken care of by the BAP layer in the 3GPP IAB architecture.

\subsection{Protocol Stack}
Figure \ref{fig:stacks} shows the protocol stack of the proposed IAB architecture for two-hop relaying (corresponding to the network in Fig. \ref{subfig:part3}). We see another difference between our architecture and the 3GPP IAB architecture --  MT has the complete UE stack instead of the partial UE stack as in the latter case. Having complete UE functionalities simplify many UE-related procedures for an MT. The figure also illustrates that the architecture uses existing/standard interfaces and protocol layers, making them economical and easier to develop and maintain. They also facilitate installing IAB-nodes anywhere without requiring additional interface modules at available cell sites, that is, they can connect to any gNB.

It may seem that an IAB-node needs more processing in our architecture than in the 3GPP counterpart due to an additional layer (i.e., PDCP + SDAP vs. BAP). However, the 3GPP solution would need packet data convergence protocol (PDCP) layer at IAB-nodes anyway for security protection of the F1 interface \cite{ts38300}. Also, when VMR and multi-RAT capabilities are to be supported in future releases, implementing the BAP layer may become more challenging than it already is.

The two-hop IAB architecture can easily accommodate multiple intermediate IAB-nodes to support multihop relaying operations. For example, Fig. \ref{subfig:multi_stacks} depicts the protocol stack for a $3$-hop relaying case. Note that the access IAB-node (IAB-node2), i.e., the serving BS for UE, has a different protocol stack than that of intermediate IAB-node (IAB-node1). By employing a full gNB Uu stack, IAB-node1 configures and maintains NR connection facing its downstream MT (of IAB-node2) as if it were a UE. The MT then uses its backhaul data radio bearer (DRB) to send C- and U-plane messages to the 5GC or communicate to the IAB-donor. Similarly, the UE communicates to the 5GC over path IAB-node2$\rightarrow$IAB-node1$\rightarrow$gNB$\rightarrow$dUPF$\rightarrow$IAB-donor-CU for both its C- and U-plane messages. 

Of course, intermediate IAB-nodes and anchor gNBs can have their own UEs. How does then IAB-node1 differentiate whether a child node is MT or UE? After registration, like a UE usually does, an MT asks the 5GC to reach its endpoint (the IAB-donor-CU) over IP. The 5GC, in response, tells the concerned gNB-CU (coincidently the IAB-donor-CU) to establish a data path to the dUPF. Because the IAB-donor-CU has been already using the dUPF to exchange messages with the MT (via gNB and the IAB-node1), it becomes aware that the UE context is actually from an MT. Next, it communicates this information to IAB-node1 over the F1 interface. The IAB-node1 marks the UE and MT contexts so that the UE data is forwarded via the F1 interface but the MT data via PDCP+SDAP layers. How does a gNB differentiate between MTs and UEs? It is straightforward -- IAB-donor is contacted by MTs only. Consequently, the IAB-node1 and gNB can treat MTs and UEs differently and manage their traffic based on certain precedence rules. Similarly, IAB-donor receives F1 packets only from the access IAB-node (and not from intermediate IAB-nodes), making it aware of UE's location within the IAB topology.

\textbf{Bearer mapping and QoS support: } Like a UE, each MT can establish multiple backhaul DRBs supporting diverse QoS needs. We notice that the C-plane signaling of a UE is sent via the U-plane of the IAB-node. To guarantee its QoS, we can reserve a DRB only to carry C-plane signaling at IAB-nodes and assign it the highest priority among DRBs. For U-plane data, an IAB-node selects an appropriate DRB to meet the QoS requirements of the data flow. Further, the IAB-node can multiplex several UE bearers (may belong to different UEs) onto a single IAB-node DRB on a similar QoS profile, typically termed many-to-one bearer mapping. Whenever the existing IAB-node bearer can no longer satisfy the QoS requirements, the IAB-donor-CU reprograms the related QoS parameters for the corresponding IAB-node bearer and sends the updated configuration to the intermediate IAB-nodes as well.

\textbf{Mobile IAB and VMR: } The proposed architecture can handle MT mobility by leveraging its full UE stack and performing the standard handover procedures. This ensures that MT mobility is transparent to the traditional network elements (and other UEs/MTs) and does not depend on whether the target BS is IAB-donor, IAB-node, or gNB. Therefore, a handover between IAB-nodes is the same as the standard inter-gNB-DU handover. In contrast, a handover between a gNB and IAB-node is actually an inter-gNB-CU handover as each IAB-node is a DU of the IAB-donor-CU. As an outcome, our architecture would experience significantly lower MT handovers and signaling overhead than the 3GPP architecture (if future releases support mobile IAB), specifically in urban deployments. 

Probably the most promising aspect of our architecture is that when a VMR moves from one gNB to another, onboard UEs do not experience any handover. As mentioned above, several UE bearers are carried through the end-to-end IP tunnel between VMR and the IAB-donor-CU. From the 5G system perspective, the UEs are still connected to the same gNB (i.e., VMR + IAB-donor-CU). In other words, there is no impact on UEs' connectivity to 5GC as neither its gNB-DU nor gNB-CU changes. Similarly, when an IAB-node performs handover between parent nodes belonging to different anchor gNBs, no handover signaling would occur for its child nodes (UEs and MTs). In the existing 3GPP solution, however, the onboard UEs will have RRC connections with the anchor BS (i.e., IAB-donor) itself. Therefore, simultaneous handovers will occur if the VMR moves across different anchor BSs (or IAB-donors), which may lead to prohibitive signaling overhead.

\textbf{Topology Adaptation and Multi-connectivity: } We can reuse existing F1AP functionality to support route redundancy in the C-plane. For example, an MT can send C-plane signaling on each backhaul link and perform RRC connection re-establishment only when all serving links become unavailable. Alternatively, MT can simultaneously connect to multiple parent nodes and manage any addition/removal of redundant routes after considering propagation conditions and traffic load on each backhaul link. Currently, 5G only supports NR-NR dual connectivity (NR-DC) at the PDCP level. Researchers are now working on extensions of NR-DC to multi-connectivity. We can additionally provide multi-connectivity at the IP level by having multiple MT instances at an IAB-node.  

\section{Multi-RAT Extension of the Architecture}
\begin{figure}[h]
\begin{minipage}{0.5\textwidth}
\begin{subfigure}{\linewidth}
\includegraphics[width=\linewidth]{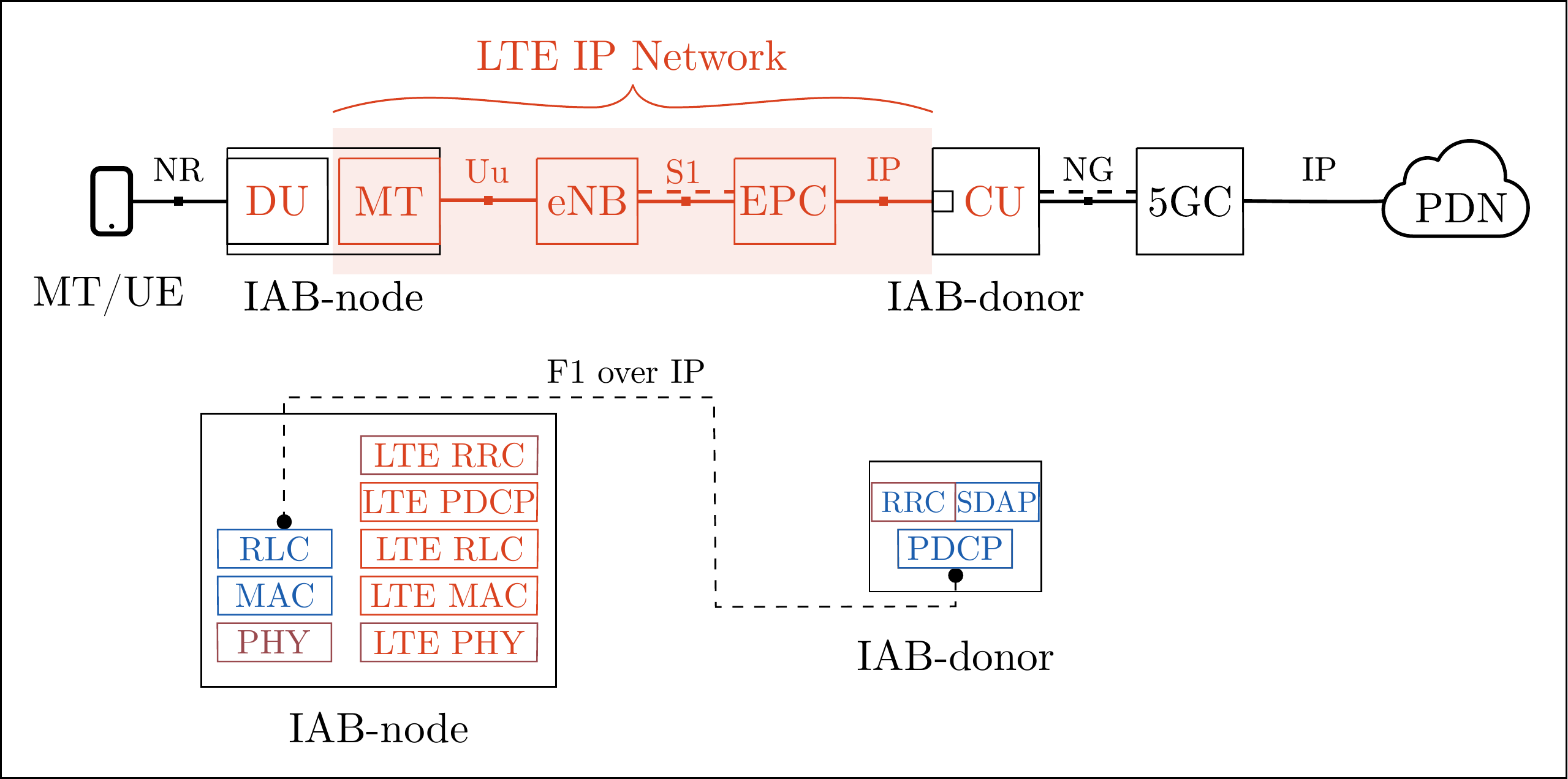}
\caption{} \label{subfig:5g_over_lte}
\end{subfigure}
\vfill
\begin{subfigure}{\linewidth}
\includegraphics[width=\linewidth]{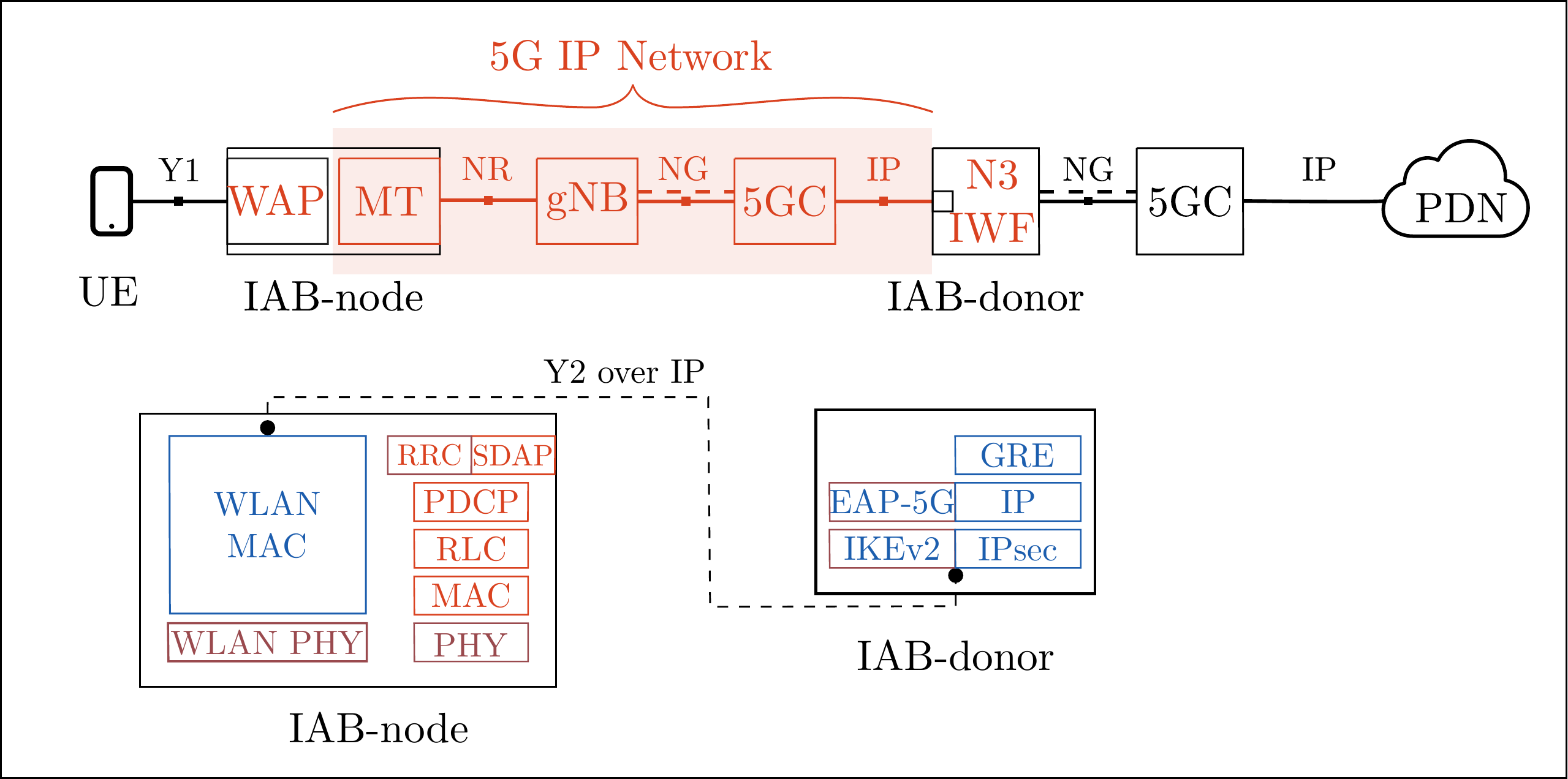}
\caption{} \label{subfig:wifi_over_5g}
\end{subfigure}
\vfill
\begin{subfigure}{\linewidth}
\includegraphics[width=\linewidth]{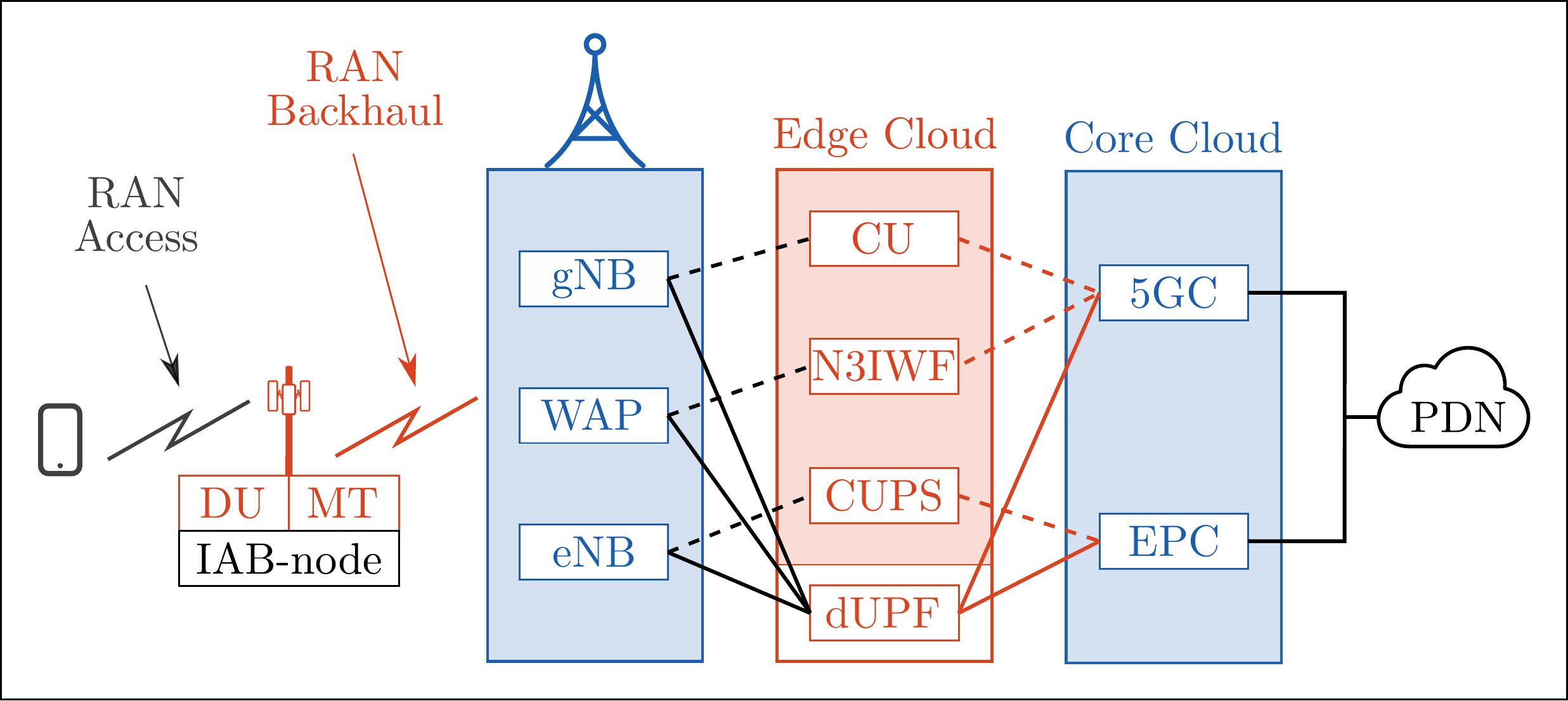}
\caption{} \label{subfig:hybrid_arch}
\end{subfigure}
\end{minipage}
\caption{Examples of some of the multi-RAT scenarios that are possible to realize by extending the proposed IAB architecture. These scenarios are (a) 5G access over LTE backhaul, (b) WLAN access over 5G backhaul, and (c) Hybrid architecture with NR, LTE, and WLAN in the RAN, and 5GC and EPC in the core network. Any combination of RATs can be used for access and backhaul.}
\label{fig:multi_rat}
\end{figure}

The proposed architecture is flexible enough to allow multi-RAT coexistence where access and backhaul can belong to different RATs (5G, LTE, or WLAN) without the need for inter-operability interfaces. For example, to provide NR access over LTE backhaul, a UE connects to IAB-node over NR interface, but the IAB-node connects to its parent node (an eNB in such case) over LTE backhaul. This necessitates that the IAB-node should possess a gNB-DU radio stack towards the UE and an LTE UE radio stack towards the eNB as depicted in Fig. \ref{subfig:5g_over_lte}. The IAB-node then establishes an IP tunnel to the IAB-donor and provides 5GC connectivity to the UE as described earlier. Consequently, a connection is created between NR UE and 5GC over an LTE IP network. Similarly, LTE access over 5G backhaul is possible if the IAB-node has an LTE eNB radio stack towards UE and an NR UE radio stack towards gNB, and the LTE network supports control and user plane separation (CUPS) mechanism \cite{ts23214}. Figure \ref{subfig:wifi_over_5g} shows the protocol stack of an IAB-node if we want to realize WLAN access (or any non-3GPP access for that matter) over 5G backhaul. The WLAN access point (WAP) part of the IAB-node establishes a Y2 interface to a non-3GPP interworking function (N3IWF) \cite{ts23501} over 5G IP network. These new envisaged deployments will support UE mobility across different RATs if the UE has radio interfaces of those RATs. 

Note that 5G access over LTE/WLAN backhaul may not provide requisite QoS to some of the DRBs established for time-sensitive or data-intensive services. This deployment is meant to be used only when anticipated services can run on LTE/WLAN infrastructures. A logical extension of multi-RAT capability is to enable dual- and multi-connectivity where UEs/MTs have multiple access links that may belong to different RATs, for example, E-UTRA-NR dual connectivity. Interestingly, it is widely accepted that the initial rollout of 5G is non-standalone (NSA) deployment, which means UEs would use NR cells for U-plane and LTE for C-plane operations. Many operators believe that NSA deployment can provide enhanced mobile broadband with reliable connectivity. Through the multi-connectivity feature, our architecture can expedite the rollout of NSA deployments before deploying 5GC and allow users to take advantage of 5G technology sooner. 

Lastly, we can design a multi-RAT multihop IAB system where a UE would have the flexibility to utilize a particular RAT, for example, based on service requirements and RAT load level. Such a hybrid cellular system is beneficial when a specific network (5G, LTE or WLAN) is partially available or unavailable (e.g., due to natural disaster) and an on-demand network is needed. A crucial aspect for realizing this, as shown in Fig. \ref{subfig:hybrid_arch}, is to have both core networks (i.e., evolved packet core (EPC) and 5GC) simultaneously in the system. We can then readily set up a relay where a core network can become independent of RAN.

\section{Performance Analysis}
To illustrate how the proposed architecture performs against the 3GPP architecture, we assess their performance on two aspects: 1) average and cell-edge throughput, and 2) mobility robustness. We evaluate these metrics using the 3GPP channel model \cite{tr36873} for the mmWave spectrum and the NYU model \cite{akdeniz2014millimeter} for beamforming gains. We assume heterogeneous deployment with $7$ hexagonal macro cell sites with inter-site distance (ISD) of $500$ m. Each cell site has $4$ outdoor small cells (IAB-nodes) installed randomly within its coverage. Each cell site also has one open-space office with dimensions $50\text{~m}\times120\text{~m}\times 3$ m that is covered by $4$ ceiling mounted indoor small cells (IAB-nodes) with ISD of $20$ m. There are $N$ UEs randomly dropped within each macrocell, of which almost half are indoor. Each BS has uniform rectangular antenna arrays with $256$ elements at both transmitter and receiver sides, whereas each UE has only one element. The rest of the system-level evaluation parameters (as recommended by \cite{tr38874, ts38214}) are summarized in Table \ref{tab:parameters}. 

\begin{table}[h]
 \scriptsize
\begin{tabular}{|p{0.34\linewidth}|p{0.56\linewidth}|}
\hline
Parameters & Values \\\hline
Carrier frequency & $30$ GHz \\\hline
System bandwidth & $1$ GHz \\\hline
Subcarrier spacing & $15$ KHz \\\hline
Thermal noise density & $-174$ dBm/Hz \\ \hline
Pathloss model & Macro to MT/UE: 5GCM UMa \newline
                 Outdoor IAB-node to MT/UE: UMi Street Canyon \newline
                 Indoor IAB-node to MT/UE: InH\\\hline 
Antenna height & Macro: $25$ m, ~Outdoor IAB-node: $10$ m, \newline 
                 Indoor IAB-node: $3$ m, ~UE: $1.5$ m \\\hline
Transmit power & Macro: $40$ dBm, ~Outdoor IAB-node: $33$ dBm, \newline
                 Indoor IAB-node: $23$ dBm, ~UE: $23$ dBm \\\hline
Noise margin & BS: $7$ dB, ~UE: $10$ dB \\\hline
Outdoor-to-indoor \newline penetration loss & $50\%$ low loss model, $50\%$ high loss model\\\hline 
MCS index table & based on \cite{ts38214}  \\\hline
\end{tabular} 
\caption{Simulation parameters for IAB evaluations.} \label{tab:parameters}
\end{table}

We consider tree topology where each MT/UE has only one parent node. A UE/MT connects to the parent node having the strongest signal-to-noise ratio (SNR) in access links. We focus on in-band backhauling and use a static time division multiplexing scheme for coordination between parent and child BSs -- in each timeslot, either odd-hop BSs (including IAB-donors) transmit and even-hop BSs receive or vice versa. Subsequently, each BS performs round-robin algorithm to serve downlink traffic to its child nodes. We assume that control signalings are instantaneous and do not occupy radio resources. Finally, Monte Carlo simulations are performed with $20$ independent runs where each one lasts for $10^6$ timeslots. 

\subsection{Average and Cell-edge Throughput} 
For the proposed architecture, MTs can connect to any gNB as they have a complete UE protocol stack. For the 3GPP architecture, we assume that only $N_d$ gNBs are deployed as IAB-donors. We use a full buffer traffic model where the network continuously sends downlink data for each UE. 

\begin{figure}[h]
\centering
\begin{subfigure}{\linewidth}
 \includegraphics[width=\linewidth]{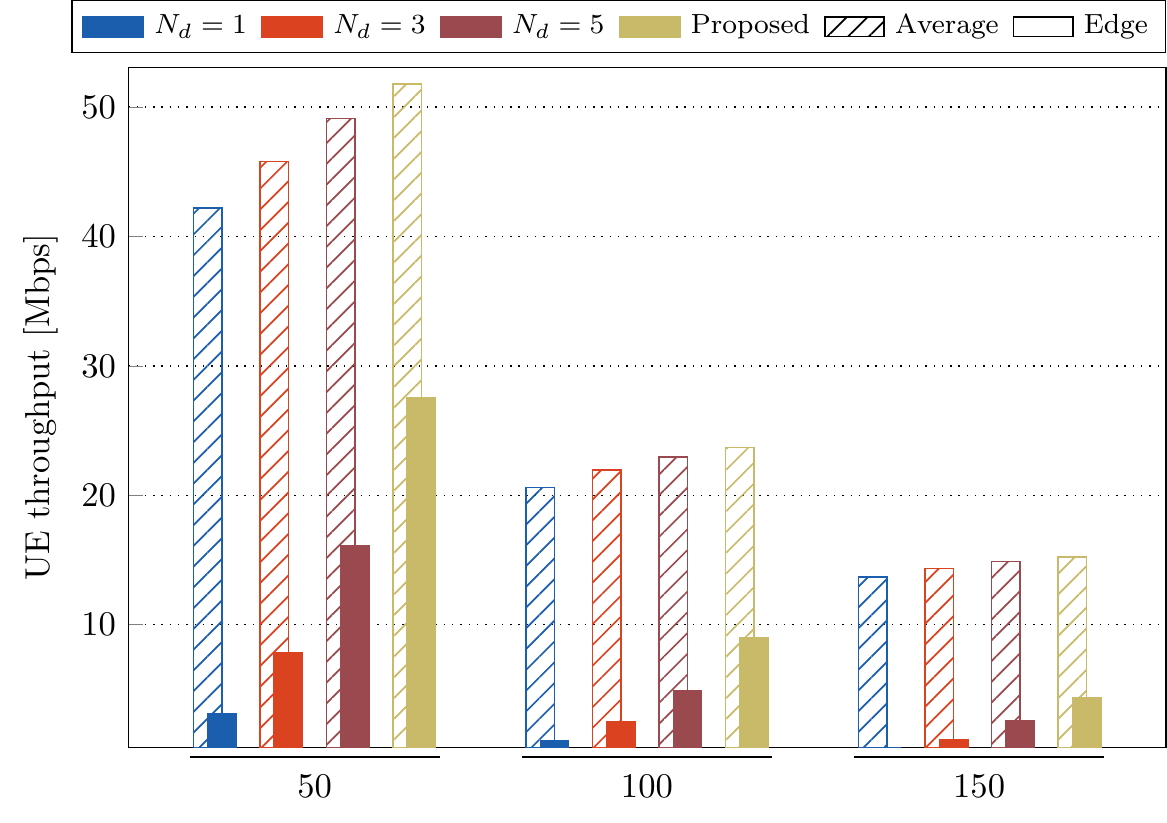}
  \caption{}\label{fig:throughput}
\end{subfigure}
\hfill
\begin{subfigure}{\linewidth}
\includegraphics[width=\linewidth]{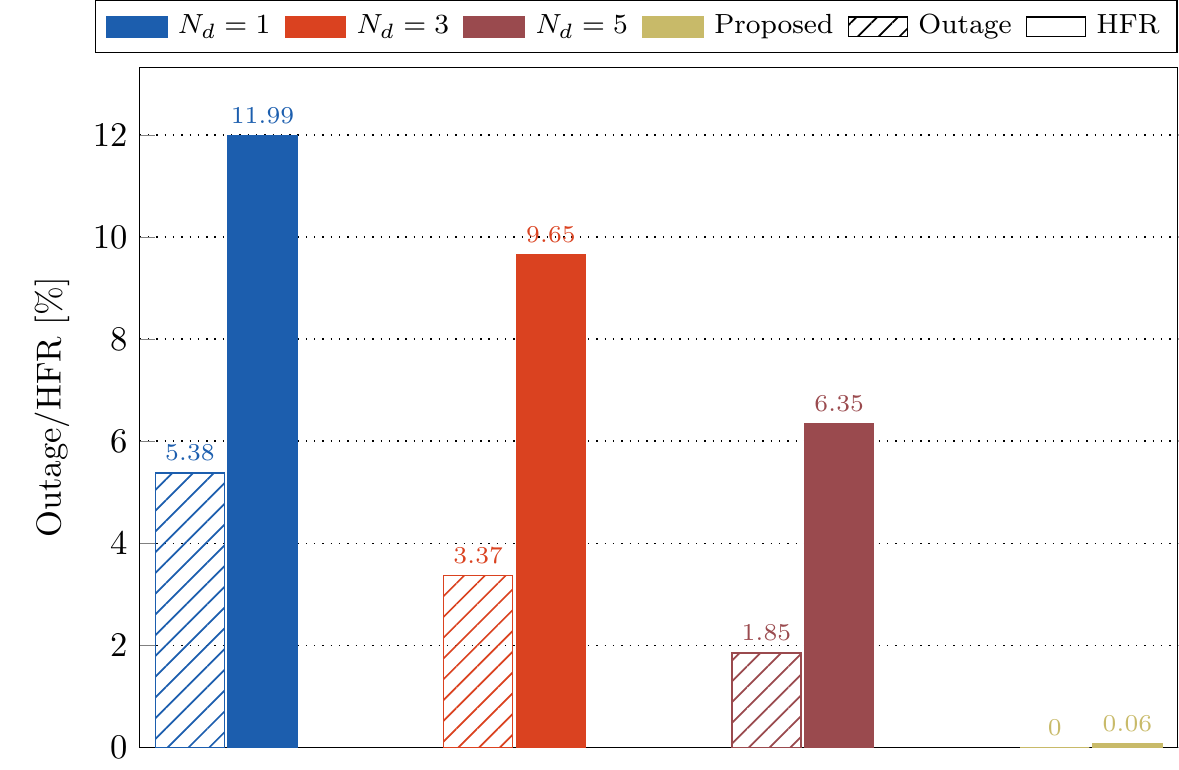}
\caption{}
\label{fig:mobility}
\end{subfigure}
 \caption{Throughput and mobility robustness comparison between the proposed architecture and the 3GPP architecture having $N_d$ IAB-donors. The plots shows (a) Average and Cell-edge throughput with different UE density per macrocell and  (b) Average outage rate and HFR when an MT is moving at $120$ kmph.} \label{fig:results}
\end{figure}

Figure \ref{fig:throughput} compares the average and cell-edge throughputs for the two architectures. The cell-edge throughput represents the $5$th percentile of user throughputs. We observe that the proposed architecture exhibits around $2-20$ percent better average UE throughput than that of the 3GPP architecture. Further, cell-edge throughput is significantly higher for the proposed architecture. In fact, we found atleast $70$ percent improvement in cell-edge throughput compared to the $N_d=5$ case. The figure also confirms the notion that with increasing $N_d$, cell-edge capacity increases multifold. 

\subsection{Mobility Robustness}
In order to assess mobility robustness, a small cell (IAB-node) is mounted on a moving bus. We assume that the bus is $3$ m high, and its transmitting and receiving antennas have $64$ elements. The bus moves according to the random waypoint mobility model within the deployment coverage at a constant speed of $120$ kmph. We further assume that beamforming vectors between the bus relay or VMR and BSs are perfectly aligned by using position estimation techniques. We do not consider the Doppler effect in this work.

The NR handover procedure used in our simulation is as follows. The VMR periodically measures SNR from serving and neighbor cells and initiates a handover process when a target cell becomes $3$ dB better than the serving cell and maintains it for $80$ ms. The serving cell then handover the VMR to the target cell if the handover preparation and execution phases complete successfully. A handover fails, however, when an RLF occurs during the handover process. The VMR then begins RLF recovery, and it takes $100$ ms to re-establish a link. The VMR declares RLF when the SNR falls below $2$ dB and does not exceed $5$ dB within $1000$ ms. We assume handover preparation delay for inter-gNB-DU handover as $20$ ms and inter-gNB-CU handover as $40$ ms, and handover execution delay for inter-gNB-DU and inter-gNB-CU handover as $25$ and $50$ ms, respectively.

We demonstrate VMR's mobility robustness through outage rate (i.e., the percentage of time it was in an outage) and handover failure ratio or HFR (i.e., the percentage of handovers that failed). The VMR is in an outage when performing RLF recovery, or SNR of its active connection is below $2$ dB. Figure \ref{fig:mobility} demonstrates that the VMR experiences no outage conditions and negligible HFR in the proposed architecture. In contrast, the 3GPP architecture indicates a much higher outage rate and HFR values. Like before, when there are fewer IAB-donors, the VMR's mobility robustness may degrade to unacceptable levels.

We emphasize that we obtained these results with simplified modeling. In reality, IAB-nodes would be deployed with careful network planning. Optimal cell selection and resource scheduling (like \cite{saha2019millimeter}) can provide better gains than presented above. Similarly, optimizing mobility-related parameters would improve mobility robustness further.

\section{Conclusion and Future Trends}
The 3GPP has standardized IAB in Release 16 as a promising cost-effective solution for ultra-densification in mmWave 5G networks. IAB enables multihop relaying, which makes deploying 5G networks in areas with partial wired backhaul or 5GC connectivity in a scalable and quick manner. The 3GPP IAB solution, however, suffers from some inherent limitations that may hinder making it flexible and self-organizing architecture by design. This article has proposed a novel alternative for B5G networks that mitigates these shortcomings. Our architecture supports VMR and multi-RAT coexistence (e.g., access and backhaul may belong to different RATs) that are not currently enabled by the existing 3GPP architecture. Our architecture does not modify any standard network element and is transparent to legacy operations of UEs, gNB, and 5GC. In fact, we have shown that our architecture fulfills almost every architectural design aspect related to IAB networks. Due to its robust design, the architecture can be rapidly standardized and deployed within the existing 5G system. Simulations have shown that the proposed architecture possesses significant gains in cell-edge throughput over that of the 3GPP architecture. Mobility robustness in terms of outage rate and HFR is also superior in the proposed architecture, especially when fewer IAB-donors are available. We believe these capabilities would give rise to a new paradigm of use-cases in the future and makes a case for standardization for B5G network.   

\bibliographystyle{IEEEtran}
\bibliography{mybib}

\end{document}